# Kepler frequency and moment of inertia of rotating neutron stars with chaotic magnetic field

Muhammad Lawrence PATTERSONS[1*]; Freddy Permana ZEN[1,2†]

[1] Theoretical Physics Laboratory, THEPi (Theoretical High Energy Physics) Division, Faculty of Mathematics and Natural Sciences, Institut Teknologi Bandung, Bandung 40132, West Java, Indonesia

[2] Indonesia Center for Theoretical and Mathematical Physics (ICTMP), Institut Teknologi Bandung, Bandung 40132, West Java, Indonesia

**ABSTRACT**

Rotating neutron stars (NSs) are crucial objects of study, as our understanding of them relies significantly on observational data from these rotating stars. Observations suggest that the magnetic fields of NSs range from approximately $10^{8-15}$ G. In this work, we compute the Kepler frequency and moment of inertia for rotating NSs under the influence of a chaotic magnetic field. We utilize an equation of state (EOS) incorporating nuclei in the crust and hyperons in the core, with the Hartle-Thorne formalism applied to address the rotational aspects. A magnetic field ansatz is selected, in which the magnetic field is coupled to the energy density. To examine the impact of a chaotic magnetic field on the Kepler frequency and moment of inertia, we vary the magnetic field strength. Our results indicate that an increase in magnetic field strength enhances the Kepler frequency of rotating NSs. For the moment of inertia, the effect of magnetic field variation is minimal at lower masses but becomes more pronounced as the mass exceeds $M = 0.5\ M_\odot$, where moment of inertia increases with increasing magnetic field. Furthermore, our results for the moment of inertia comply with constraint derived from pulsar mass measurements, data from gravitational wave events GW170817 and GW190425, and X-ray observations of emission from hotspots on NS surfaces measured by NICER.

Keywords: Chaotic Magnetic Field, Kepler Frequency, Moment of Inertia, Neutron Stars

## 1. INTRODUCTION

Much of our understanding of neutron stars (NSs) comes from observations of pulsars, which are rotating NSs [1]. Since the discovery of the first pulsar by Hewish et al. [2], a significant number of pulsars have been identified, and they are believed to be highly magnetized NSs [1]. Anomalous X-ray pulsars (AXPs) and soft gamma repeaters (SGRs) are recognized as potential magnetars—NSs characterized by ultra-strong magnetic fields exceeding $4.414 \times 10^{13}$ G [3]. Observational data indicate that magnetic field strengths in NSs span a broad range, from approximately $10^8$ G to $10^{15}$ G [4]. Although direct observation of the magnetic field in NS cores remains unachieved, theoretical predictions suggest field strengths on the order of $10^{18}$ G to $10^{20}$ G [5].

Theoretical model for rotating NSs has been formulated by Hartle & Thorne [6]. In the Hartle-Thorne (HT) formalism, rotation is considered as a perturbation to the static configuration. Applying this formalism to NS calculations enables the determination of physical properties to the NS mass and radius. Using this formalism, one can calculate the total mass, radius, and angular momentum of NSs. HT formalism has been widely applied in studies of rotating NSs (see Refs. [7–13]). Furthermore, angular velocity as a rotational aspect of NSs is constrained by a limit known as the Kepler frequency, which depends on the NS's mass and radius in a static configuration [1, 10, 11]. It is worth noting that the authors in Refs. [10, 11] conducted calculations of the Kepler frequency within their respective works.

Another important property of NSs is their moment of inertia, a key physical quantity that provides insights into the internal structure of the stars, particularly their equation of state (EOS). Given the connection between the EOS and moment of inertia, it is possible to develop approximations for estimating the moment of inertia, even within static models, especially when considering stiff EOS cases [14]. Moreover, a measurement of NS moment inertia is crucial due to its universal relationship with the star's compactness [15].

---

[*] m.pattersons@proton.me

[†] fpzen@fi.itb.ac.id

In term of magnetic field impact on NSs, several studies have investigated the impact of magnetic fields on NSs. Konno et al. [16] examined the deformation of NSs with a polytropic equation of state (EOS). Mallick & Schramm [17] focused on mass correction and deformation in magnetized NSs, while Lopes & Menezes [18] explored the effects of chaotic magnetic fields on NSs by introducing an ansatz for magnetic fields coupled to energy density. Franzon et al. [19] studied the internal composition of NSs and proto NSs under strong magnetic field. Pattersons et al. [20] investigated the impact of chaotic magnetic field on mass-radius relation of rotating NSs.

In this work, we calculate the Kepler frequency and moment of inertia for rotating NSs with chaotic magnetic fields, employing the HT formalism as the rotational framework and the Lopes-Menezes ansatz to model the magnetic field. We use the EOS proposed by Miyatsu et al. [21], which includes nuclei in the crust and hyperons in the core. The numerical simulation algorithm follows that used in Refs. [10] and [22].

The structure of this paper is as follows: Section 2 describes the formalisms used in our analysis. In Section 3, we present the numerical results along with a discussion of their significance. Finally, Section 4 provides a summary of our key findings.

## 2. MATHEMATICAL FORMALISMS

To provide a self-contained explanation, Subsection 2.1 provides a brief review of the chaotic magnetic field of neutron stars, while Subsection 2.2 covers the HT formalism. It is important to note that in the mathematical formulations, we adopt the geometrized units $G = c = 1$, where $G$ denotes the universal gravitational constant, and $c$ is the speed of light.

### 2.1 Chaotic Magnetic Field

Magnetic field could form anisotropy of the NSs [23]. According to Ref. [18], if the magnetic field is in $z$-direction, the stress tensor writes: $diag\left(\frac{B^2}{8\pi}; \frac{B^2}{8\pi}; -\frac{B^2}{8\pi}\right)$, being non identical. There exists an argument that the effect of a magnetic field can be described using the concept of pressure only in the case of a small-scale chaotic field. Under this condition, the pressure due to the magnetic field $p_B$ is shown to be consistent with field theory [18, 20, 24]. By agreeing this argument, $p_B$ writes

$$p_B = \frac{1}{3}\langle T_a^a \rangle = \frac{1}{3}\left(\frac{B^2}{8\pi} + \frac{B^2}{8\pi} - \frac{B^2}{8\pi}\right) = \frac{B^2}{24\pi}, \tag{1}$$

where $T_a^a$ denotes the spatial components of energy-momentum tensor (EMT). Since now the pressures in all direction are equal to $p_B$, the anisotropy vanishes. Now we can write the total energy density $\rho$ and the total pressure $p$ as

$$\rho = \rho_m + \frac{B^2}{8\pi}, \tag{2}$$

$$p = p_m + \frac{B^2}{24\pi}, \tag{3}$$

where the subscript $m$ denotes the matter contribution.

In this work, we use the Lopes-Menezes ansatz of magnetic field, which reads [18, 20]

$$B = B_0 \left(\frac{\rho_m}{\rho_0}\right)^\psi + B_s, \tag{4}$$

where $B_0$ can be interpreted as the magnetic field at the center of NS, $\rho_0$ is the energy density at the center of the NS with maximum mass when the magnetic field is zero, $\psi$ is any positive number, and $B_s$ denotes the magnetic field at the surface of the star. It has to be noted that in this work we employ $\psi = 0.001$. For the argument supporting this choice, refer to Ref. [20]. It is important to note that in this study, we vary the magnetic field strength, with the specific variations detailed in Table 1.

Table 1. Variations of the magnetic field

| Set Name | $B_0$ | $B_s$ |
|:---:|:---:|:---:|
| Set 1 | 0 | 0 |
| Set 2 | $1 \times 10^{18}$ G | $10^9$ G |

| Set Name | $B_0$ | $B_s$ |
|---|---|---|
| Set 3 | $2 \times 10^{18}$ G | $10^{12}$ G |
| Set 4 | $3 \times 10^{18}$ G | $10^{15}$ G |

## 2.2 Hartle-Thorne Formalism

In HT approximation of rotational configuration, the metric reads [6–8]
$$ds^2 = -e^{2\nu} dt^2 + e^{2\lambda} dr^2 + r^2 e^{2\gamma}(d\phi - \omega\, dt)^2 + r^2 e^{2\xi} d\theta^2, \tag{5}$$
where $\omega$ is the angular velocity of the local inertial frame which is generated by frame-dragging. We can expand exponentials factor in Eq. (5) to be
$$e^{2\nu} = e^{2\varphi}\left[1 + 2(h_0 + h_2 P_2(\cos\theta))\right], \tag{6}$$
$$e^{2\lambda} = \left[1 + \frac{2}{r}(m_0 + m_2 P_2(\cos\theta))\left(1 - \frac{2m(r)}{r}\right)^{-1}\right], \tag{7}$$
$$e^{2\gamma} = \sin^2\theta\left[1 + 2(v_2 - h_2)P_2(\cos\theta)\right], \tag{8}$$
$$e^{2\xi} = \left[1 + 2(v_2 - h_2)P_2(\cos\theta)\right], \tag{9}$$
where $h_0, h_2, m_0, m_2$, dan $v_2$ are the perturbation functions; $P_2(\cos\theta)$ denotes the second order of Legendre polynomial function; $e^{2\varphi}$ dan $m(r)$ are the functions constrained by
$$\frac{d\varphi}{dr} = \frac{m(r) + 4\pi r^3 p(r)}{r(r - 2m(r))}, \tag{10}$$
$$\frac{dm}{dr} = 4\pi r^2 \rho(p), \tag{11}$$
It is important to note that calculating the Kepler frequency and moment of inertia only requires solving the unperturbed terms of the metric. Therefore, in this study, we do not address expansion terms of the metric.

Pressure is calculated using Tolman-Oppenheimer-Volkoff (TOV) equation, which writes [25]
$$\frac{dp}{dr} = -\frac{(\rho + p)\, m(r) + 4\pi r^3 p(r)}{r(r - 2m(r))}. \tag{12}$$
Here we have $p(R) = 0$, where $R$ is the radius of NS. Mass-radius relation of NS can be obtained by solving simultaneously Eqs. (10), (11), and (12).

Considering the rotational configuration, the Einstein field equation $R^t_\phi = 8\pi T^t_\phi$ would give [26]
$$\frac{1}{r^4}\frac{d}{dr}\left(r^4 j \frac{d\bar{\omega}}{dr}\right) + \frac{4}{r}\frac{dj}{dr}\bar{\omega} = 0, \tag{13}$$
where $\bar{\omega}$ denotes the angular velocity of NS relative to the local inertial frame which satisfies $\bar{\omega} = \Omega - \omega$. Here $\Omega$ is the angular velocity of the star relative to the distant observers where the frame-dragging vanishes. In this work, we choose $\Omega = 1000$ s$^{-1}$. In Eq. (13), $j$ is given by
$$j = e^{-\varphi}\sqrt{1 - \frac{2m(r)}{r}}. \tag{14}$$
The boundary condition for Eq. (13) is $\bar{\omega}(r = 0) = \omega_c$, where $\omega_c$ can be chosen arbitrarily. In this work, we choose $\omega_c = 80$ s$^{-1}$. Refer to Ref. [20] for the reasoning of this choice.

By calculating the Einstein field equation for the zeroth order expansion terms, we obtain [6–8, 26]
$$\frac{dm_0}{dr} = 4\pi r^2 \frac{d\epsilon}{dp}(\epsilon + p)p_0^* + \frac{1}{12}j^2 r^4\left(\frac{d\bar{\omega}}{dr}\right)^2 - \frac{1}{3}r^3\frac{dj^2}{dr}\bar{\omega}^2, \tag{15}$$

$$\frac{dp_0^*}{dr} = -\frac{m_0(1 + 8\pi r^2 p)}{(r - 2m)^2} - \frac{4\pi(\epsilon + p)r^2}{r - 2m}p_0^* + \frac{1}{12}\frac{r^4 j^2}{r - 2m}\left(\frac{d\bar{\omega}}{dr}\right)^2 + \frac{1}{3}\frac{d}{dr}\left(\frac{r^3 j^2 \bar{\omega}}{r - 2m}\right), \tag{16}$$
where $p_0^*$ is the pressure perturbation factor. It is important to note that Eq. (15) and Eq. (16) are coupled differential equations, with the solution for $m_0$ being central to both. The boundary conditions at $r = 0$ for are $m_0 = 0$ and $p_0^* = 0$. The mass correction due to rotation is associated to the value of $m_0$ at $r = R$, where $R$ is the radius of the star. Mathematically, the mass correction writes
$$\Delta M = m_0(R) + \frac{L^2}{R^3}. \tag{17}$$

Here $L$ denotes the angular momentum of the NSs, which reads [10]

$$L(\Omega) = \frac{8\pi}{3} \int_0^R r^4 \frac{\rho + p}{\sqrt{1 - \frac{2m}{r}}} \bar{\omega} e^{-\varphi} \, dr \quad. \tag{18}$$

The total mass of rotating NSs is given by

$$M = M_0 + \Delta M \quad. \tag{19}$$

where $M_0$ is mass of the star obtained by solving Eq. (11).

From Eq. (18), we can calculate the moment of inertia

$$I = \frac{L}{\Omega} \quad. \tag{20}$$

Finally, using the approximation presented in Ref. [1], the Kepler frequency of a NS is given by [10, 11]

$$\Omega_K = 0.65 \left(\frac{M_0}{R^3}\right)^{\frac{1}{2}} \quad. \tag{21}$$

## 3. RESULTS AND DISCUSSION

This section presents the findings from our numerical calculations, highlighting key results and their implications within the context of Kepler frequency and moment of inertia of NSs. We solve all differential equations using Euler method. It should be noted that the mass plotted in the figures within this section represents the total mass in the rotational configuration. For details on the relationship between rotational mass and radius of NSs with the same parameters used here, please refer to Ref. [20].

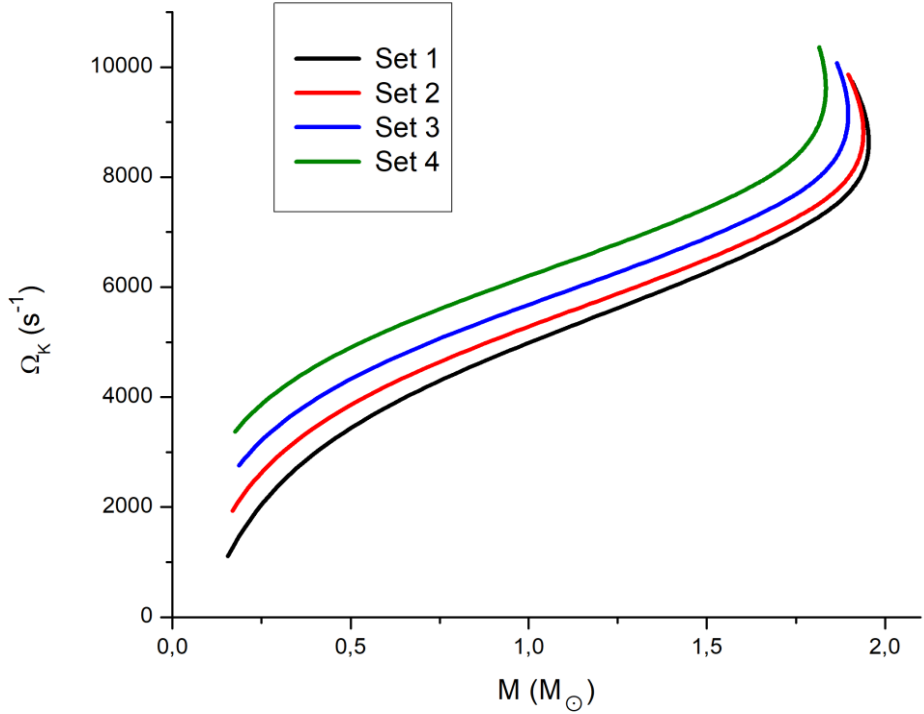

Figure 1. Relation between the mass and Kepler frequency of rotating NSs

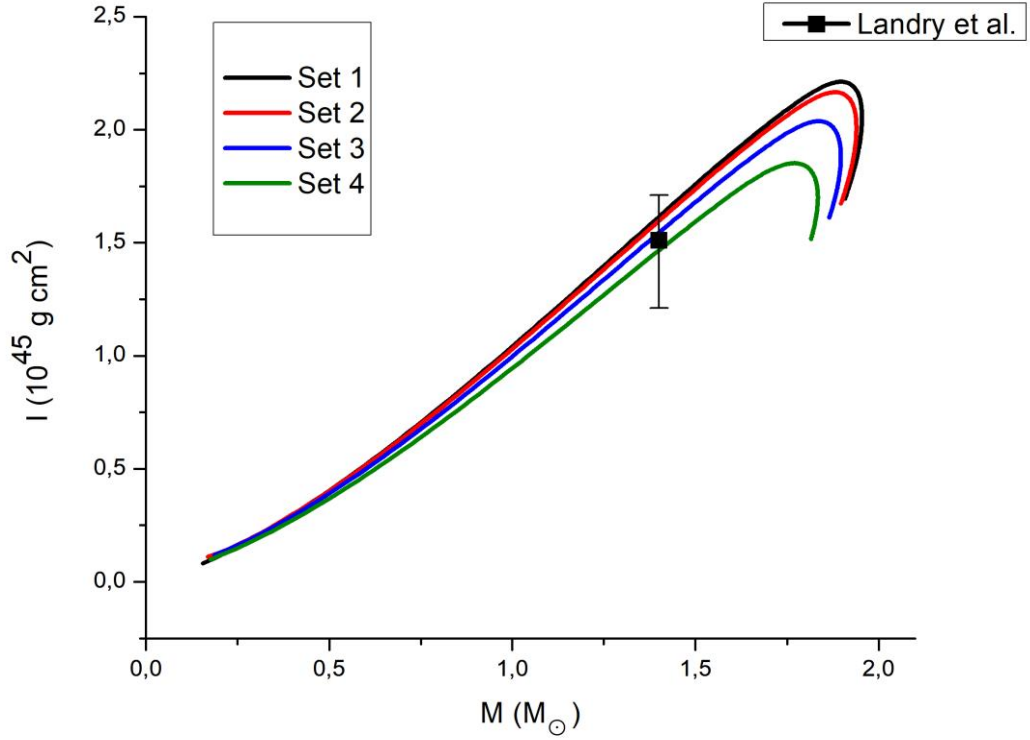

Figure 2. Relation between the mass and moment of inertia of rotating NSs

Fig. 1 shows the relation between the rotational mass and Kepler frequency of NSs. It can be inferred that the Kepler frequency increases with chaotic magnetic field strength, suggesting that NSs with stronger chaotic magnetic fields can rotate faster than the angular velocity limit of those with weaker fields. Notably, Ref. [20] indicates that the formation of rotating NSs with stronger chaotic magnetic fields leads to lower maximum masses and reduced radii. Our findings are consistent with angular momentum conservation, where the formation of NSs with stronger magnetic fields exhibit lower maximum mass, reduced maximum radius, and increased maximum angular velocity (i.e., Kepler frequency).

In Fig. 2, the relation between the rotational mass and moment of inertia of NSs is illustrated. At lower masses, the curves coincide, indicating that in this regime, the influence of chaotic magnetic field strength on the moment of inertia is minimal. A clear separation of the curves begins at $M = 0.5\ M_\odot$, where a stronger chaotic magnetic field reduces the moment of inertia. Since the moment of inertia is proportional to both the mass and the square of the radius, these findings align with the mass-radius relation of NSs presented in Ref. [20].

Interestingly, our result for the moment of inertia falls within the constraint range presented by Landry et al. [27]. This constraint aims to verify whether rotating NSs with chaotic magnetic fields are compatible with observational data. The constraint value is $I = 1.51^{+020}_{-0.30} \times 10^{45}$ g cm$^2$ at $M = 1.4\ M_\odot$, the canonical mass of NSs. It is worth noting that observations indicate that NSs typically occupy a relatively narrow mass range near 1.4 $M_\odot$ [28]. According to Ref. [27]. the constraint derived from pulsar mass measurements, data from gravitational wave events GW170817 and GW190425, and X-ray observations of emission from hotspots on NS surfaces measured by NICER.

## 4. CONCLUSION

In this study, we explore the impact of chaotic magnetic fields coupled to energy density on the Kepler frequency and moment of inertia of rotating NSs. Our findings show that the chaotic magnetic field increases the Kepler frequency of rotating NSs, while, conversely, it reduces their moment of inertia. Moreover, the moment of inertia of rotating NSs with chaotic magnetic fields falls within the constraint range derived from pulsar mass measurements, data from gravitational wave events (GW170817 and GW190425), and X-ray observations of emission from hotspots on NS surfaces measured by NICER.

## ACKNOWLEDGEMENTS

MLP sincerely thanks Anna Campoy Ordaz for making her code publicly available, which served as the basis for the code used in this work. MLP also gratefully acknowledges financial support from the Indonesia Endowment Fund for Education (LPDP). FPZ would like to thank Kemendikbudristek Republic of Indonesia (through LPPM ITB, Indonesia) for partially financial support.